\documentclass[aoas,nameyear,MSNbibl,rotating,dvips]{arximspdf}
\usepackage{graphicx}

\doi{10.1214/10-AOAS442}
\volume{5}
\issue{2B}
\pubyear{2011}
\firstpage{1379}
\lastpage{1406}

\makeatletter

\let\sv@tabnotetext\tabnotetext
  \let\sv@tabnotemark@fmt\tabnotemark@fmt
   \long\def\legend#1{{\let\tabnote@indent\leavevmode\sv@tabnotetext[]{}{#1}}}

\def\bptnote#1{}

\def\@bmisc[#1]{%
  \get@battribute{unstr}%
  \common@pub@types%
  \let\bauthor\bbl@bauthor%
  \let\bhowpublished\@firstofone%
  \def\borganization##1{{\bauthor@style ##1}}%
}

\newcommand{\dm}[1]{#1}

\makeatother

\begin{document}
\begin{frontmatter}

\title{Forecasting emergency medical service call arrival rates\thanksref{TITL}}
\pdftitle{Forecasting emergency medical service call arrival rates}
\runtitle{Forecasting EMS call arrival rates}
\thankstext{TITL}{Supported in part by NSF Grant CMMI-0926814.}

\begin{aug}
\author[A]{\fnms{David S.} \snm{Matteson}\corref{}\ead[label=e1]{dm484@cornell.edu}},
\author[A]{\fnms{Mathew W.} \snm{McLean}},
\author[A]{\fnms{Dawn B.} \snm{Woodard}}
and~\author[B]{\fnms{Shane G.} \snm{Henderson}}
\runauthor{Matteson, McLean, Woodard and Henderson}
\affiliation{Cornell University}
\address[A]{
School of Operations Research\\
\quad and Information Engineering\\
Cornell University\\
282 Rhodes Hall\\
Ithaca, New York 14853\\
USA\\
\printead{e1}} 
\end{aug}

\received{\smonth{7} \syear{2010}}
\revised{\smonth{11} \syear{2010}}

%
\begin{abstract}
We introduce a new method for forecasting emergency call arrival rates
that combines integer-valued time series models with a~dynamic latent
factor structure. Covariate information is captured via simple
constraints on the factor loadings. We directly model the count-valued
arrivals per hour, rather than using an artificial assumption of
normality. This is crucial for the emergency medical service context,
in which the volume of calls may be very low. Smoothing splines are
used in estimating the factor levels and loadings to improve long-term
forecasts. We impose time series structure at the hourly level, rather
than at the daily level, capturing the fine-scale dependence in
addition to the long-term structure.

Our analysis considers all emergency priority calls received by Toronto
EMS between January 2007 and December 2008 for which an ambulance was
dispatched. Empirical results demonstrate significantly reduced error
in forecasting call arrival volume. To quantify the impact of reduced
forecast errors, we design a queueing model simulation that
approximates the dynamics of an ambulance system. The results show
better performance as the forecasting method improves. This notion of
quantifying the operational impact of improved statistical procedures
may be of independent interest.
\end{abstract}

%

\end{frontmatter}

\section{Introduction}\label{sec:intro}

Considerable attention has been paid to the problem of how to best
deploy ambulances within a municipality to minimize their response
times to emergency calls while keeping costs low. Sophisticated
operations research models have been developed to address issues such
as the optimal number of ambulances, where to place bases, and how to
move ambulances in real time via system-status management [\citet
{swersey1994deployment}; \citet{gol04}; \citet{henderson2009}]. However, methods for
estimating the inputs to these models, such as travel times on road
networks and call arrival rates, are ad hoc. Use of inaccurate
parameter estimates in these models can result in poor deployment
decisions, leading to low performance and diminished user confidence in
the software. We introduce methods for estimating the demand for
ambulances, that is, the total number of emergency calls per period,
that are highly accurate, straightforward to implement, and have the
potential to simultaneously lower operating costs while improving
response times.

Current practice for forecasting call arrivals is often rudimentary.
For instance, to estimate the call arrival rate in a small region over
a specific time period, for example,\ next Monday from 8--9 a.m., simple
estimators have been constructed by averaging the number of calls
received in the corresponding period in four previous weeks: the
immediately previous two weeks and the current and previous weeks of
the previous year. Averages of so few data points can produce highly
noisy estimates, with resultant cost and efficiency implications.
Excessively large estimates lead to over-staffing and unnecessarily
high costs, while low estimates lead to under-staffing and slow
response times. \citet{setzler2009ems} document an emergency medical
service (EMS) agency which extends this simple moving average to twenty
previous observations: the previous four weeks from the previous five
years. 
A more formal time series approach is able to account for possible
differences from week to week and allows inclusion of neighboring hours
in the estimate.

We generate improved forecasts of the call-arrival volume by
introducing an integer-valued time series model with a dynamic latent
factor structure for the hourly call arrival rate. Day-of-week and
week-of-year effects are included via simple constraints on the factor
loadings. The factor structure allows for a significant reduction in
the number of model parameters. Further, it provides a systematic
approach to modeling the diurnal pattern observed in intraday counts.
Smoothing splines are used in estimating the factor levels and
loadings. This may introduce a small bias in some periods, but it
offers a significant reduction in long-horizon out-of-sample
forecast-error variance. This is combined with integer-valued time
series models to capture residual dependence and to provide adaptive
short-term forecasts. Our empirical results demonstrate significantly
reduced error in forecasting hourly call-arrival volume.

Few studies have focused specifically on EMS call arrival rates, and of
those that have proposed methods for time series modeling, most have
been based on Gaussian linear models. Even with a continuity
correction, this method is highly inaccurate when the call arrival rate
is low, which is typical of EMS calls at the hourly level. Further, it
conflicts with the Poisson distribution assumption used in operations
research methods for optimizing staffing levels. For example, \citet
{channouf2007application} forecast EMS demand by modeling the daily
call arrival rate as Gaussian, with fixed day-of-week, month-of-year,
special day effects and fixed day-month interactions. They also
consider a Gaussian autoregressive moving-average (ARMA) model with
seasonality and special day effects. Hourly rates are later estimated
either by adding hour-of-day effects or assigning a multinomial
distribution to the hourly volumes, conditional on the daily call
volume estimates.

\citet{setzler2009ems} provide a comparative study of EMS call volume
predictions using an artificial neural network (ANN). They forecast at
various temporal and spatial granularities with mixed results. Their
approach offered a significant improvement at low spatial granularity,
even at the hourly level. At a high spatial granularity, the mean
square forecast error (MSFE) of their approach did not improve over
simple binning methods at a temporal granularity of three hours or less.

Methods for the closely related problem of forecasting call center
demand have received much more study. \citet{bianchi1998improving} and
\citet{andrews1995ll} use ARMA models to improve forecasts for daily
call volumes in a retail company call center and a telemarketing
center, respectively. A dynamic harmonic regression model for hourly
call center demand is shown in \citet{tych2002unobserved} to outperform
seasonal ARMA models. Their approach accounts for possible
nonstationary periodicity in a time series. The major drawback common
to these studies is that the integer-valued observations are assumed to
have a continuous distribution, which is problematic during periods
with low arrival rates.

The standard industry assumption is that hourly call-arrival volume has
a Poisson distribution. The Palm--Khintchine theorem---stating that the
superposition of a number of independent point processes is
approximately Poisson---provides a theoretical basis for this
assumption [see, e.g., \citet{whi02}]. \citet{brown2005statistical}
provide a comprehensive analysis of operational data from a bank call
center and thoroughly discuss validating classical queueing theory,
including this theorem. \citet{henderson2005should} states that we can
expect the theorem to hold for typical EMS data because there are a~%
large number of callers who can call at any time and each has a very
low probability of doing so.

\citet{weinberg2007bayesian} use Bayesian techniques to fit a
nonhomogeneous Poisson process model for call arrivals to a commercial
bank's call center. This approach has the advantage that forecast
distributions for the rates and counts may be easily obtained. They
incorporate smoothness in the within-day pattern. They implement a
variance stabilizing transformation to obtain approximate normality.
This approximation is most appropriate for a Poisson process with high
arrival rates, and would not be appropriate for our application in
which very low counts are observed in many time periods. 

\citet{shen2008interday} apply the same variance stabilizing
transformation and achieve better performance than \citet
{weinberg2007bayesian}. They use a singular value decomposition (SVD)
to reduce the number of parameters in modeling arrival rates. Their
approach is used for intraday updating and forecasts up to one day
ahead. 

\citet{shen2008forecasting} propose a dynamic factor model for
15-minute call arrivals to a bank call center. They assume that call
arrivals are a Cox process. A Cox process [cf. \citet{cox1980point}] is
a Poisson process with a~stochastic intensity, that is, a doubly
stochastic Poisson process. The factor structure reduces the number of
parameters by explaining the variability in the call arrival rate with
a small number of unobserved variables. Estimation proceeds by
iterating between an SVD and fitting Poisson generalized linear models
to successively estimate the factors and their respective loadings. The
intensity functions are assumed to be serially dependent. Forecasts are
made by fitting a simple autoregressive time series model to the factor
score series.\looseness=-1

We assume that the hourly EMS call-arrival volume has a Poisson
distribution. This allows parsimonious modeling of periods with small
counts, conforms with the standard industry assumption, and avoids use
of variance stabilizing transformations. We assume the intensity
function is a random process and that it can be forecast using previous
observations. This has an interpretation very similar to a Cox process,
but is not equivalent since the random intensity is allowed to depend
on not only its own history, but also on previous observations. We
partition the random intensity function into stationary and
nonstationary components.

Section \ref{sec:discuss} describes the general problem and our data set. Section \ref{3.0}
presents the proposed methodology. We consider a dynamic latent factor
structure to model the nonstationary pattern in intraday call arrivals
and greatly reduce the number of parameters. We include day-of-week and
week-of-year covariates via simple constraints on the factor loadings
of the nonstationary pattern. Smoothing splines are easily incorporated
into estimation of the proposed model to impose a smooth evolution in
the factor levels and loadings, leading to improved long-horizon
forecast performance. We combine the factor model with stationary
integer-valued time series models to capture the remaining serial
dependence in the intensity process. This is shown to further improve
short-term forecast performance of our approach. A simple iterative
algorithm for estimating the proposed model is presented. It can be
implemented largely through existing software. Section \ref{4} assesses the
performance of our approach using statistical metrics and a queueing
model simulation. Section \ref{sec:con} gives our concluding remarks.

\section{Notation and data description}\label{sec:discuss}

We assume that over short, equal-length time intervals, for example,
one hour periods, the latent call arrival intensity function can be
well approximated as being constant, and that all data have been
aggregated in time accordingly.
We suppose aggregated call arrivals follow a nonhomogeneous counting
process $\{Y_t \dvtx t \in\mathbb{Z} \}$, with discrete time index $t$.
Underlying this is a latent, real-valued, nonnegative intensity
process $\{\lambda_t \dvtx t \in\mathbb{Z} \}$. We further assume that
conditional on $\lambda_t$, $Y_t$ has a~Poisson distribution with mean
$\lambda_t$.

\begin{figure}

\includegraphics{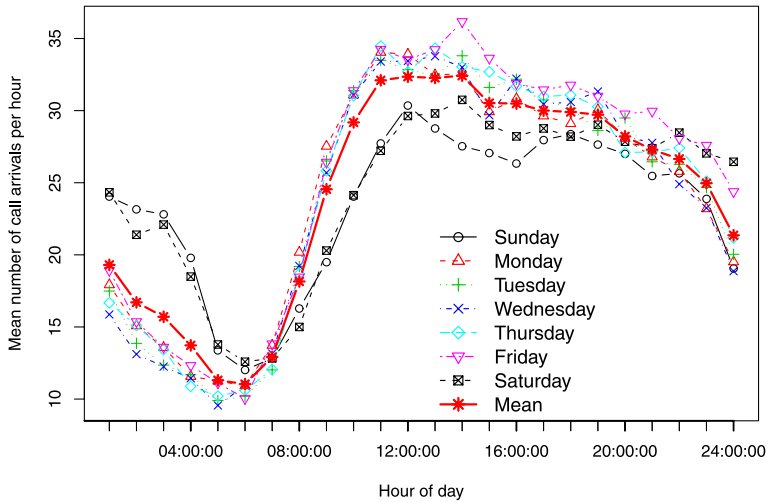}

\caption{Mean number of calls per hour by day of the week.}
\label{calls}
\end{figure}

As shown in Figure \ref{calls}, the pattern of call arrivals over the
course of a~typical day has a distinct shape. After quickly increasing
in the late morning, it peaks in the early afternoon, then slowly falls
until it troughs between 5~and 6 a.m. See Section~\ref{4} for more
discussion. In our analysis, we consider an arrival process that has
been repeatedly observed over a particular time span, specifically, a
24 hour day. Let
\[
\{y_t \dvtx t = 1,\ldots,n\} \equiv\{y_{ij} \dvtx i = 1,\ldots,d; j =
1,\ldots,m\}
\]
denote the sequence of call arrival counts, observed over time period
$t$, which corresponds one-to-one with the $j$th sub-period of the
$i$th day, so that $n = dm$. Our baseline approach is to model the
arrival intensity $\lambda_{t}$ for the distinct shape of intraday
call arrivals using a small number of smooth curves. 

We consider two disjoint information sets for predictive conditioning.
Let $\mathcal{F}_t = \sigma(Y_1, \ldots, Y_t)$ denote the $\sigma
$-field generated by $Y_1, \ldots, Y_t$, and let $\mathbf{X} = \{
\mathbf{x}_1, \ldots, \mathbf{x}_n\}$ denote any available
deterministic covariate information about each observation. We
incorporate calendar information such as day-of-week and week-of-year
in our analysis. We define $\lambda_t$ as the conditional expectation
of $Y_t$ given $\mathcal{F}_{t-1}$ \textit{and} $\mathbf{X}$.
We defined this above as the mean of $Y_t$. In our \textit{model} these
coincide; however, this mean may not be the same as the conditional
expectation since $\lambda_t$ may depend on other unobserved random variables.
Let $\mu_t = E(Y_t |\mathbf{X} ) > 0$ denote the conditional mean of
$Y_t$ given only the covaria\-tes~$\mathbf{X}$. Let
%
\begin{equation}
\lambda_t
= E(Y_t | \mathcal{F}_{t-1} , \mathbf{X} )
= \mu_t E ( {Y_t}/{\mu_t} | \mathcal{F}_{t-1} , \mathbf{X} )
= \mu_t \eta_t,
\end{equation}
in which $\eta_t > 0$ is referred to as the conditional intensity
inflation rate (CIIR). By construction, 
\[
E(\eta_t | \mathbf{X} )
= E(E ( {Y_t}| \mathcal{F}_{t-1} , \mathbf{X} ) | \mathbf{X} ) /{\mu_t}
= E( {Y_t} | \mathbf{X} ) /{\mu_t}
= 1.
\]

The CIIR process is intended to model any remaining serial dependence
in call arrival counts after accounting for available covariates. In
the EMS context we hypothesize that this dependence is due to sporadic
events such as inclement weather or unusual traffic patterns. Since
information regarding these events may not be available or predictable
in general, we argue that an approach such as ours which explicitly
models the remaining serial dependence will lead to improved short-term
forecast accuracy. \dm{In Section \ref{3.0} we consider a dynamic
latent factor model estimated with smoothing splines for modeling $\mu
_t$, various time series models for modeling $\eta_t$, and finally a~%
conditional likelihood algorithm for estimating the latent intensity
process~$\lambda_t$ via estimation of $\eta_t$ given $\mu_t$.}

The call arrival data used consists of all emergency priority calls
received by Toronto EMS between January 1, 2007 and December 31, 2008
for which an ambulance was dispatched. This includes some calls not
requiring lights-and-sirens response, but does not include scheduled
patient transfers.
We include only the first call arrival time in our analysis when
multiple calls are received for the same event. The data were processed
to exclude calls with no reported location. These removals totaled less
than 1\% of the data.

Many calls resulted in multiple ambulances being dispatched.
Exploratory analysis revealed that the number of ambulances deployed
for a single emergency did not depend on the day of the week, the week
of the year, or exhibit any serial dependence. However, such instances
were slightly more prevalent in the morning hours.
Our analysis of hourly ambulance demand defines an event as a call
arrival if \textit{one or more} ambulances are deployed. 

We removed seven days from the analysis because there were large gaps,
over at least two consecutive hours, in which no emergency calls were
received. These days most likely resulted from malfunctions in the
computer-aided dispatch system which led to failures in recording calls
for extended periods. Strictly speaking, it is not necessary to remove
the entire days; however, we did so since it had a negligible impact on
our results and it greatly simplified out-of-sample forecast
comparisons and implementation of the simulation studies in Section
\ref{4}.

Finally, we gave special consideration to holidays. We found that the
intraday pattern on New Year's Eve and Day was fundamentally different
from the rest of the year and removed these days from our analysis.
This finding is similar to the conclusions of \citet
{channouf2007application} who found that New Year's Day and the dates
of the Calgary Stampede were the only days requiring special
consideration in their methodology when applied to the city of Calgary.
In practice, staffing decisions for holidays require special planning
and consideration of many additional variables. 

\section{Modeling}\label{3.0}

Factor models provide a parsimonious representation of high dimensional
data in many applied sciences, for example, econometrics [cf. \citet
{stock2002macroeconomic}]. We combine a dynamic latent factor model
with integer-valued time series models. We include covariates via
simple constraints on the factor loadings. We estimate the model using
smoothing splines to impose smooth evolution in the factor levels and
loadings. The factor model provides a parsimonious representation of
the nonstationary pattern in intraday call arrivals, while the time
series models capture the remaining serial dependence in the arrival
rate process.

\subsection{Dynamic latent factor model}\label{3.1}

For notational simplicity, assume $m$ consecutive observations per day
are available for $d$ consecutive days with no omissions in the record.
Let $\mathbf{Y} = (y_{ij})$ denote the $d \times m$ matrix of observed
counts for each day $i$ over each sub-period $j$.
Let $\mu_{ij} = E(Y_{ij} | {\mathbf{X}} )$, and let $\mathbf{M} =
(\mu_{ij})$ denote the corresponding $d \times m$ latent nonstationary
intensity matrix.
To reduce the dimension of the intensity matrix $\mathbf{M}$, we
introduce a~$K$-factor model.

We assume that the intraday pattern of expected hourly call arrivals on
the log scale can be well approximated by a linear combination of (a
small number) $K$ factors or functions, denoted by $\mathbf{f}_{k}$
for $k = 1,\ldots,K$. The factors are orthogonal length-$m$ vectors. The
intraday arrival rate model $\bolds{\mu}_{i}$ over a~particular day $i$
is given by
%
\begin{equation}\label{m1.5}
\log\bolds{\mu}_{i}
= {L_{i1}} {\mathbf{f}_{1}} + \cdots+ {L_{iK}} {\mathbf{f}_{K}}.
\end{equation}
Each of the factors $\mathbf{f}_{k}$ varies as a function over the
periods within a day, but they are constant from one day to the next.
Day-to-day changes are modeled by allowing the various factor loadings
$L_{i k}$ to vary across days.
When $K$ is much smaller than either $m$ or $d$, the dimensionality of
the general problem is greatly reduced. In practice, $K$ must be chosen
by the practitioner; we provide some discussion on choosing $K$ in
Section \ref{4}.

In matrix form we have
%
\begin{equation}\label{m1}
\log\mathbf{M}
= \mathbf{L} \mathbf{F}^{\mathrm{T}},
\end{equation}
in which $\mathbf{F} = (\mathbf{f}_1, \ldots, \mathbf{f}_K)$
denotes the $m \times K$ matrix of underlying factors and $\mathbf{L}$
denotes the corresponding $d \times K$ matrix of factor loadings, both
of which are assumed to have full column rank. Although other link
functions are available, the component-wise log transformation implies
a multiplicative structure among the $K$ common factors and ensures a
positive estimate of each hourly intensity $\mu_{ij}$. Since neither
$\mathbf{F}$ nor $\mathbf{L}$ are observable, the expression (\ref
{m1}) is not identifiable. We further require $\mathbf{F}^{\mathrm{T}}
\mathbf{F} = \mathbf{I}$ to alleviate this ambiguity and we
iteratively estimate $\mathbf{F}$ and $\mathbf{L}$.

\subsection{Factor modeling with covariates via constraints}\label{3.2}

To further reduce the dimensionality, we impose a set of constraints on
the factor loading matrix $\mathbf{L}$. Let $\mathbf{H}$ denote a $d
\times r$ full rank matrix ($r<d$) of given constraints (we discuss
later what these should be for EMS). Let $\mathbf{B}$ denote an $r
\times K$ matrix of unconstrained factor loadings. These unconstrained
loadings $\mathbf{B}$ linearly combine to constitute the constrained
factor loadings $\mathbf{L}$, such that $\mathbf{L} = \mathbf{H}
\mathbf{B}$. Our factor model may now be written as
\[
\log\mathbf{M}
= \mathbf{L} \mathbf{F}^{\mathrm{T}}
= \mathbf{H} \mathbf{B} \mathbf{F}^{\mathrm{T}}.
\]
A considerable reduction in dimensionality occurs when $r$ is much
smaller than $d$.

Constraints to assure identifiability are standard in factor analysis.
The constraints we now consider incorporate auxiliary information about
the rows and columns of the observation matrix $\mathbf{Y}$ to
simplify estimation and to improve out-of-sample predictions. Similar
constraints have been used in \citet{takane2001constrained}, \citet
{tsaitsay09} and \citet{mattesontsay10}.

For example,
the rows of $\mathbf{H}$ might consist of incidence vectors for
particular days of the week, or special days which might require unique
loadings on the common factors. We may choose to constrain all weekdays
to have identical factor loadings and similarly constrain weekend days.
However, this approach is much more general than simple equality
constraints, as demonstrated below.

The intraday pattern of hourly call arrivals varies from one day to the
next, although the same general shape is maintained. As seen in Figure
\ref{calls}, different days of the week exhibit distinct patterns. We
do not observe large changes from one week to the next, but there are
significant changes over the course of the year. We allow loadings to
slowly vary from week to week. Both of these features are incorporated
into the factor loadings $\mathbf{L}$ by specifying appropriate
constraints $\mathbf{H}$. Let
%
\begin{equation} \label{m4}
\log\mathbf{M}
= \mathbf{L} \mathbf{F}^{\mathrm{T}}
= \mathbf{H} \mathbf{B} \mathbf{F}^{\mathrm{T}}
=
\pmatrix{
\mathbf{H}^{(1)} & \mathbf{H}^{(2)}
}
\pmatrix{
\mathbf{B}^{(1)} \cr
\mathbf{B}^{(2)}
}
\mathbf{F}^{\mathrm{T}},
\end{equation}
in which the first term corresponds to day-of-week effects and the
second to smoothly varying week-of-year effects. $\mathbf{H}^{(1)}$ is
a $d \times7$ matrix in which each row $\mathbf{H}^{(1)}_i$ is an
incidence vector for the day-of-week. Similarly, $\mathbf{H}^{(2)}$ is
a $d \times53$ matrix in which each row $\mathbf{H}^{(2)}_i$ is an
incidence vector for the week-of-year. \dm{(We use a 53 week year
since the first and last weeks may have fewer than 7 days.)} The $7
\times K$ matrix $\mathbf{B}^{(1)} = (\mathbf{b}^{(1)}_1, \ldots,
\mathbf{b}^{(1)}_K)$ contains unconstrained factor loadings for the
day-of-week and $\mathbf{B}^{(2)} = (\mathbf{b}^{(2)}_1, \ldots,
\mathbf{b}^{(2)}_K)$ is a $53 \times K$ matrix of factor loadings for
the week-of-year.

\subsection{Factor model estimation via smoothing splines}\label{3.3}

We assume that as the nonstationary intensity process $\mu_{ij}$
varies over the hours $j$ of each day $i$, it does so smoothly. If each
of the common factors $\mathbf{f}_{k} \in\mathbb{R}^m$ varies
smoothly over sub-periods $j$, then the smoothness of $\mu_{ij}$ is
guaranteed for each day. Increasing the number of factors reduces
possible discontinuities between the end of one day and the beginning
of the next. To incorporate smoothness into the model (\ref{m1.5}), we
use Generalized Additive Models (GAMs) in the estimation of the common
factors $\mathbf{f}_k$. GAMs extend generalized linear models,
allowing for more complicated relationships between the response and
predictors by modeling some predictors nonparametrically [see, e.g.,
\citet{hastie-generalized}; \citet{wood2006generalized}]. GAMs have been
successfully used for count-valued data in the study of fish
populations [cf. \citet{borchers1997improving}; \citet{daskalov1999relating}].
%
%
The factors $\mathbf{f}_k = {f}_k(j)$ are a smooth function of the
intraday time index covariate $j$. The loadings $\mathbf{L}$ are
defined as before. If the loadings $\mathbf{L}$ were known covariates,
equation (\ref{m1.5}) would be a varying coefficient model [cf. \citet
{hastie1993varying}].


There are several excellent libraries available in the statistical
package \textit{R} [\citet{Rsoftware}] for fitting GAMs, thus making them
quite easy to implement. We used the \textit{gam} function from the \textit{mgcv} library [\citet{Rmgcv}] extensively. Other popular libraries
include the \textit{gam} package [\citet{Rgam}] and the \textit{gss} package
[\citet{Rgss}]. See \citeauthor{wood2006generalized} [(\citeyear{wood2006generalized}), Section~5.6] for an
introduction to GAM estimation using \textit{R}.

In estimation of the model (\ref{m1.5}) via the \textit{gam} function, we
have used thin plate regression splines with a ten-dimensional basis,
the Poisson family, and the log-link function. Thin plate regression
splines are a low rank, isotropic smoother with many desirable
properties. For example, no decisions on the placement of knots is
needed. They are an optimal approximation to thin plate splines and,
with the use of Lanczos iteration, they can be fit quickly even for
large data sets [cf. \citet{wood2003thin}].

When the factors $\mathbf{F}$ are treated as a fixed covariate, the
factor model can again be interpreted as a varying coefficient model.
Given the calendar covariates $\mathbf{X}$, let
%
\begin{eqnarray}\label{m33}
\log{\mu}_{ij}&=& {F}_{j1} L^{\mathrm{T}}_{1i} + \cdots+ {F}_{jK} L^{\mathrm
{T}}_{Ki} \nonumber\\
&=& \sum_{k = 1}^K {F}_{jk} \bigl\{ {\mathbf{H}^{(1) \mathrm{T}}_i} \mathbf
{b}_k^{(1)} + \mathbf{H}^{(2) \mathrm{T}}_i \mathbf{b}_k^{(2)} \bigr\}\\
&=& \sum_{k = 1}^K {F}_{jk} \bigl\{ {b}_k^{(1)}(\mathbf{x}_i) +
{b}_k^{(2)}(\mathbf{x}_i) \bigr\},\nonumber
\end{eqnarray}
in which ${b}_k^{(1)}(\mathbf{x}_i)$ is a piece-wise constant function
of the day-of-week, and ${b}_k^{(2)}(\mathbf{x}_i)$ is a smoothly
varying function over the week-of-year. We may again proceed with
estimation via the \textit{gam} function in \textit{R}. Day-of-week
covariates are simply added to the linear predictor as indicator
variables. These represent a level shift in the daily loadings on each
of the factors $\mathbf{f}_{k}$. In our application it is appropriate
to assume a smooth transition between the last week of one year and the
first week of the next. To ensure this in estimation of
${b}_k^{(2)}(\mathbf{x}_i)$, we use a cyclic cubic regression spline
for the basis [cf. \citet{wood2006generalized}, Section 5.1]. Iterative
estimation of $\mathbf{F}$, and $\mathbf{L}$ via $\mathbf{B}$, for a
given number of factors $K$ is discussed in Section \ref{3.5}.

We allow the degree of smoothness for the factors ${f}_{k}$ and the
loadings function ${b}_k^{(2)}(\mathbf{x}_i)$ to be automatically
estimated by generalized cross validation (GVC). We expect short term
serial dependence in the residuals for our application. For smoothing
methods in general, if autocorrelation between the residuals is
ignored, automatic smoothing parameter selection may break down [see,
e.g., \citet{Opsomer01nonparametricregression}]. The proposed factor
model may be susceptible to this if the number of days included is not
sufficiently large compared to the number of smooth factors and
loadings, or if the residuals are long-range dependent.
We use what is referred to as a \textit{performance} iteration [cf.
\citet{gu92}] versus an \textit{outer} iteration strategy which requires
repeated estimation for many trial sets of the smoothing parameters.
The performance iteration strategy is much more computationally
efficient for use in the proposed algorithm, but convergence is not
guaranteed, in general. In particular, \textit{cycling} between pairs of
smoothing parameters and coefficient estimates may occur [cf. \citet
{wood2006generalized}, Section 4.5], especially when the number of
factors $K$ is large. 

\subsection{Adaptive forecasting with time series models}\label{3.4}

Let $\widehat{e}_t = Y_{t}/\widehat{\mu}_{t}$ denote the
multiplicative residual in period $t$ implied by the fitted values
$\widehat{\mu}_{t}$ from a factor model estimated as described in the
previous sections. Time series plots of this residual process appear
stationary, but exhibit some serial dependence. In this section we
consider time series models for the latent CIIR process $\eta_t = E (
{Y_t}/{\mu_t} | \mathcal{F}_{t-1} , \mathbf{X} )$ to account for
this dependence.

\begin{figure}

\includegraphics{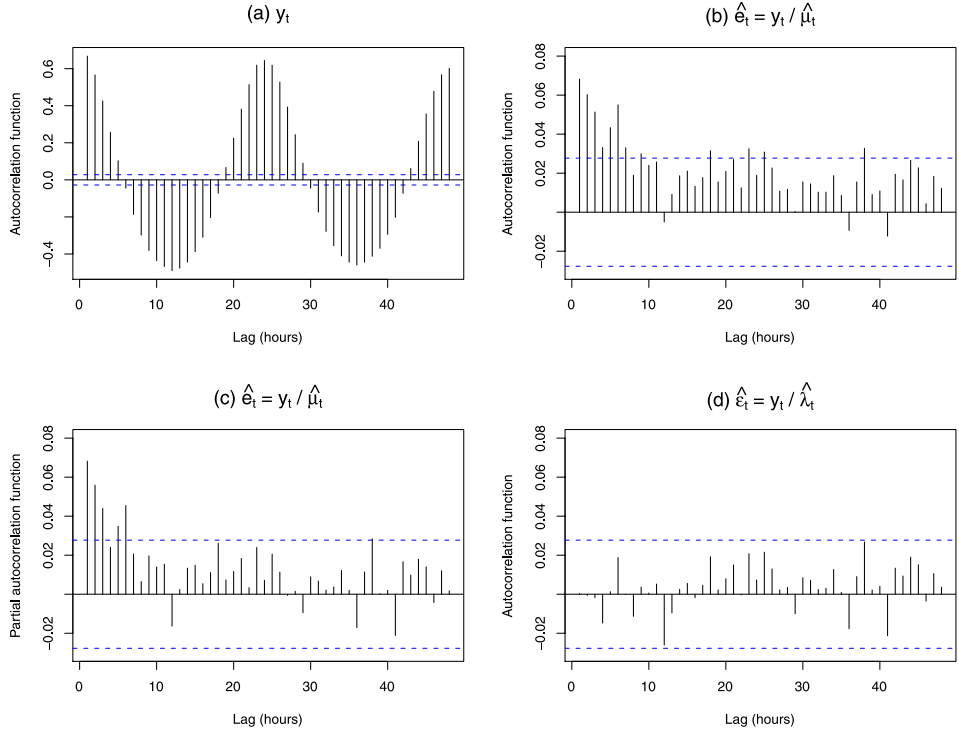}

\caption{\textup{(a)} Sample autocorrelation function for
hourly call arrival counts $y_t$. Residual $\widehat{e}_t =
y_t/\widehat{\mu}_t$ \textup{(b)} autocorrelation and \textup{(c)} partial
autocorrelation functions for fitted factor model $\widehat{\mu}_t$
with $k = 4$ factors using factor and loading constraints and smoothing
splines. \textup{(d)}~Standardized residual $\widehat{\varepsilon}_t =
y_t/\widehat{\lambda}_t = y_t/(\widehat{\mu}_t\widehat{\eta}_t)$
autocorrelation function for fitted factor model with fitted
$\operatorname{IntGARCH}(1,1)$ model for $\eta_t$. Dashed lines give approximate 95\%
confidence levels.} \label{ACF}
\end{figure}

To investigate the nature of the serial dependence, we study the
bivariate relationship between the $\widehat{e}_t$ process versus
several lagged values of the process $\widehat{e}_{t-\ell}$.
Scatterplots reveal a roughly linear relationship. Residual
autocorrelation and partial autocorrelation plots for one of the factor
models fit in Section \ref{4} are given in Figure \ref{ACF}(b) and (c).
These quantify the strength of the linear relationship as the lag $\ell
$ increases. It appears to persist for many periods, with an
approximately geometric rate of decay as the lag increases.

To explain this serial dependence, we first consider a generalized
autoregressive linear model, defined by the recursion
%
\begin{equation}\label{intgarch}
\eta_t = \omega+ \alpha\widehat{e}_{t-1} + \beta\eta_{t-1}.
\end{equation}
To ensure positivity, we restrict $\omega> 0$ and $\alpha, \beta\ge
0$. When $\mu_{t}$ is constant, the resulting model for $Y_t$ is an
$\operatorname{Integer\mbox{-}GARCH}(1,1)$ (IntGARCH) model [e.g., \citet{ferlatora06}]. It is
worth noting some properties of this model for the constant $\mu_{t}$
case. To ensure the stationarity of $\eta_t$, we further require that
$\alpha+ \beta< 1$. This sum determines the persistence of the
process, with larger values of $\alpha$ leading to more adaptability.
When this stationarity condition is satisfied, and $\eta_t$ has
reached its stationary distribution, the expectation of $\eta_t$ given
${\mathbf{X}}$ is
\begin{eqnarray*}
E( \eta_t | {\mathbf{X}}) = \omega/ ( 1 - \alpha- \beta).
\end{eqnarray*}
To ensure $E(\eta_t | {\mathbf{X}} ) = 1$ for the fitted model, we
may parameterize $\omega= 1 - \alpha- \beta$. This constraint is
simple enough to enforce for the model (\ref{intgarch}) and we do so.
However, additional parameter constraints such as this may make
numerical estimation intractable in more complicated models and they
are not enforced by us in the models outlined below.

When $\mu_{t}$ is a nonstationary process, the conditional intensity
\[
\lambda_t = \mu_{t}\eta_t
\]
is also nonstationary. Since $E(\eta_t | {\mathbf{X}} ) = 1$, we
interpret $\eta_t$ as the stationary multiplicative deviation, or
inflation rate, between $\lambda_t$ and $\mu_t$. The $\lambda_t$
process is mean reverting to the $\mu_t$ process. Let
\[
\widehat{\varepsilon}_t = Y_{t}/\widehat{\lambda}_{t}
\]
denote the multiplicative \textit{standardized} residual process given an
estimated CIIR process $\widehat{\eta}_t$. If a fitted model defined
by (\ref{intgarch}) sufficiently explains the observed linear
dependence in $\widehat{e}_t $, then an autocorrelation plot of
$\widehat{\varepsilon}_t$ should be statistically insignificant for
all lags $\ell$. As a preview, the standardized residual
autocorrelation plot for one such model fit in Section \ref{4} is
given in Figure \ref{ACF}(d). The serial correlation appears to have
been adequately removed.

\dm{
Next, we formulate three different nonlinear generalizations of (\ref
{intgarch}) that may better characterize the serial dependence, and
possibly lead to improved forecasts. The first is an exponential
autoregressive model defined as
%
\begin{eqnarray}
\eta_t & = & \alpha\widehat{e}_{t-1} + [\beta+ \delta\exp(-\gamma
\eta_{t-1}^2)] \eta_{t-1},
\end{eqnarray}
in which $\alpha, \beta, \delta, \gamma> 0$. Exponential
autoregressive models are attractive in application because of their
threshold-like behavior. For large $\eta_{t-1}$, the functional
coefficient for $\eta_{t-1}$ is approximately $\beta$, and for small
$\eta_{t-1}$ it is approximately $\beta+ \delta$. Additionally, the
transition between these regimes remains smooth. As in \citet{fokrahtjo09},
for $\alpha+ \beta< 1$ one can verify the $\eta_t$ process has a
stationarity version when $\mu_{t}$ is constant.
}


\dm{
We also consider a piecewise linear threshold model
%
\begin{equation} \label{thresholdmodel}
\eta_t  =  \omega+ \alpha\widehat{e}_{t-1} + \beta\eta_{t-1}
+ ( \gamma\widehat{e}_{t-1} + \delta\eta_{t-1} )
I_{\{\widehat{e}_{t-1} \notin(c_1,c_2)\}},
\end{equation}
in which $I$ is an indicator variable and the threshold boundaries
satisfy $0 < c_1 < 1 < c_2 < \infty$. To ensure positivity of $\eta
_t$, we assume $\omega, \alpha, \beta> 0$, $(\alpha+ \gamma) > 0$,
and $(\beta+ \delta) > 0$. Additionally, we take $\delta\leq0$ and
$\gamma\geq0$, such that when $\widehat{e}_{t-1}$ is outside the
range $(c_1,c_2)$ the CIIR process $\eta_t$ is more adaptive, that is,
puts more weight on $\widehat{e}_{t-1}$ and less on $\eta_{t-1}$.
When $\mu_{t}$ is constant, the $\eta_t$ process has a stationary
version under the restriction $\alpha+ \beta+ \gamma+ \delta< 1$;
see \citet{woomathen10}. In practice, the threshold boundaries $c_1$
and $c_2$ are fixed during estimation, and may be adjusted as necessary
after further exploratory analysis. We chose $c_1 = 1/1.15$ and $c_2 =
1.15$, that is, thresholds at 15\% above and below 1.

Finally, we consider a model with regime switching at deterministic
times, letting
%
\begin{equation} \label{regime}
\qquad \eta_t  =  (\omega_1 + \alpha_1 \widehat{e}_{t-1} + \beta_1 \eta
_{t-1} )
I_{\{t \in(t_1,t_2)\}}
+ ( \omega_2 + \alpha_2 \widehat{e}_{t-1} + \beta_2 \eta_{t-1} )
I_{\{t \notin(t_1,t_2)\}}.
\end{equation}
This model is appropriate assuming the residual process has two
distinct regimes for different periods of the day. For example, one
regime could be for normal workday hours with the other regime being
for the evening and early morning hours. No stationarity is possible
for this model. A drawback of this model is that the process has jumps
at $t_1$ and $t_2$. As was the case for $c_1$ and $c_2$ in (\ref
{thresholdmodel}), $t_1$ and $t_2$ are fixed during estimation. After
exploratory analysis, we chose $t_1=10$ a.m. and $t_2=4$ p.m.
}

\subsection{Estimation algorithm}\label{3.5}

The estimation procedure below begins with an iterative algorithm for
estimating the factor model from Sections \ref{3.1}--\ref{3.3}
through repeated use of the \textit{gam} function from the \textit{mgcv}
library in \textit{R}. Any serial dependence is ignored during estimation
of $\mu_{t}$ for simplicity. Given estimates for the factor model
$\widehat{\mu}_{t}$, conditional maximum likelihood is used to
estimate the conditional intensity $\lambda_{t}$ via one of the time
series models given in (\ref{intgarch})--(\ref{regime}) for the CIIR
process $\eta_{t}$.

\begin{enumerate}

\item Initialization:
\begin{enumerate}[(a)]
\item[(a)] Fix $K$ and $\mathbf{H}$.
\item[(b)] Choose some $c \in(0,1)$ and define $\mathbf{Y}_c = (y_{ij}
\vee c)$.
\item[(c)] Apply a singular value decomposition (SVD) to find $\log(
\mathbf{Y}_c ) =\break \mathbf{U}_0\mathbf{D}_0\mathbf{V}^{\mathrm{T}}_0$.
\begin{enumerate}[(iii)]
\item[(i)] Let $\mathbf{U}_0^{(1:K)}$ denote the first $K$ columns of the
left singular matrix $\mathbf{U}_0$.
\item[(ii)] Let $\mathbf{V}_0^{(1:K)}$ denote the first $K$ columns of the
right singular matrix $\mathbf{V}_0$.
\item[(iii)] Let $\mathbf{D}_0^{(1:K)}$ denote the upper-left $K \times K$
sub-matrix of $\mathbf{D}_0$, the diagonal matrix of singular values.
\end{enumerate}
\item[(d)] Assign $\mathbf{L}_0 = \mathbf{U}_0^{(1:K)}\mathbf
{D}_0^{(1:K)}$ and $\mathbf{F}_0 = \mathbf{V}_0^{(1:K)}$.
\item[] No smoothing is performed and the constraints $\mathbf{H}$
are omitted in initialization.
\end{enumerate}

\item Update:

\begin{enumerate}[(a)]
\item[(a)] Fit the Poisson GAM model described in Section \ref{3.3} with
$\mathbf{F}= \mathbf{F}_n$ and $\mathbf{H}$ as fixed covariates.
\begin{itemize}
\item Assign $\mathbf{B}_{n^*}$ as the estimated parameter values from
this fit and let $\mathbf{L}_{n^*} = \mathbf{H} \mathbf{B}_{n^*}$.
\end{itemize}

\item[(b)] Fit the Poisson GAM model described in Section \ref{3.3} with
$\mathbf{L} = \mathbf{L}_{n^*}$ as a fixed covariate.
\begin{itemize}
\item Assign $\mathbf{F}_{n^*}$ as the estimated parameter values from
this fit.
\end{itemize}

\item[(c)] Apply an SVD to find $\mathbf{B}_{n^*} \mathbf
{F}_{n^*}^{\mathrm{T}} = \mathbf{U}_{n+1}\mathbf{D}_{n+1}\mathbf
{V}^{\mathrm{T}}_{n+1}$.
\begin{enumerate}[(iii)]
\item[(i)] Assign $\mathbf{B}_{n+1} = \mathbf{U}_{n+1}^{(1:K)}\mathbf
{D}_{n+1}^{(1:K)}$.
\item[(ii)] Assign $\mathbf{F}_{n+1} = \mathbf{V}_{n+1}^{(1:K)}$.
\item[(iii)] Assign $\mathbf{L}_{n+1} = \mathbf{H} \mathbf{B}_{n+1}$.
\end{enumerate}
\item[(d)] Let $\log\mathbf{M}_{n+1} = \mathbf{L}_{n+1}\mathbf
{F}_{n+1}^{\mathrm{T}}$.
\end{enumerate}

\item Repeat the \textit{update} steps recursively until convergence.

\end{enumerate}

Convergence is reached when the relative change in $\mathbf{M}$ is
sufficiently small. After convergence we can recover $\log\widehat
{\mu_t}$ from the rows of the final estimate of $\log\mathbf{M}$.
These values are then treated as fixed constants during estimation of
$\eta_{t}$. We use conditional maximum likelihood to estimate the
parameters $(\omega, \alpha, \beta, \ldots)$ associated with a time
series model for $\eta_{t}$. The recursion defined by (\ref
{intgarch})--(\ref{regime}) requires initialization by choosing a value
for $\eta_1$; the estimates are conditional on the chosen initialization.

We may always specify the joint distribution $P_ \mathbf{Y}$ of the
observations $\mathbf{Y}$ as an iterated product of successive
conditional distributions $P_{Y_t}$ for $Y_{t}$ given $(Y_{t-1},\ldots
,Y_{1})$ as
\[
P_Y(y_T,y_{T-1},\ldots,y_2,y_1) = P_{Y_1}(y_1) \prod_{t=2}^{T}
P_{Y_t}(y_t | y_{t-1},\ldots, y_{1}).
\]
We follow the standard convention of fixing $P_{Y_1}(y_1) = 1$ in
estimation. For large sample sizes the practical impact of this
decision is negligible. We may therefore write the log likelihood
function as the sum of iterated conditional log likelihood functions.
The conditional distribution for the observations is assumed to be
Poisson with mean $\lambda_{t} = {\mu}_t \eta_t .$

For uninterrupted observations over periods $1,\ldots,T,$ we define
the log likelihood function as
%
\begin{eqnarray}\label{loglik1}
{\ell}(\omega, \alpha, \beta, \ldots| \widehat{M}, Y, \eta_1) &=&
\sum_{t=2}^{T} \ell_t (\omega, \alpha, \beta, \ldots| y_t,
y_{t-1}, \widehat{\mu}_t,\widehat{\mu}_{t-1},\eta_{t-1}) \nonumber
\\
&=& \sum_{t=2}^{T} (y_t \log\lambda_t - \lambda_t - \log y_t ! )
\\
&=& \sum_{t=2}^{T} \bigl(y_t \log( \widehat{\mu}_t \eta_t ) - \widehat
{\mu}_t \eta_t - \log y_t ! \bigr).\nonumber
\end{eqnarray}
This recursion requires an initial value for $\eta_1$. For simplicity,
we use its expected value, $\eta_1 = 1$. When there are gaps in the
observation record, equation~(\ref{loglik1}) is calculated over every
contiguous block of observations. This requires reinitialization of
$\eta_t = 1$ at the beginning of each block. The log likelihood for
the blocks are then added together to form the entire log likelihood.
The maximum likelihood estimate is the $\mathrm{argmax}$ of this
quantity, subject to the constraints given in Section \ref{3.4}.
Finally, $\eta_t$ is estimated by the respective recursion given by
equations (\ref{intgarch})--(\ref{regime}) with parameters replaced by
their estimates, again with reinitialization of $\eta_t = 1$ at the
beginning of each contiguous block of observations. Blocks were large
enough in our application that the effect of reinitialization was negligible.

\section{Empirical analysis}\label{4}

Using the data described in Section \ref{sec:discuss}, we perform the following analysis:
(a) we define various statistical goodness-of-fit metrics suitable for
the proposed models; based on in-sample performance, these metrics are
used to determine the number of factors $K$ for use in the dynamic
factor models. (b)~\dm{We compare the out-of-sample forecast
performance for the factor model in (\ref{m1}), the factor model with
constraints in (\ref{m4}), and the factor model with constraints and
smoothing splines in (\ref{m33}). These comparisons help ascertain the
improvement from each refinement and validate the proposed selection
methods for $K$. (c) For the latter factor model,} we compare the
out-of-sample forecast performance with the addition of the CIIR
process, via use of the various time series models defined in Section
\ref{3.4}. (d) We quantify the practical impact of these successive
statistical improvements with a queueing application constructed to
approximate ambulance operations.

\subsection{Interpreting the fitted model}\label{4.1}

The mean number of calls was approximately 24 per hour for 2007 and
2008, and no increasing or decreasing linear trend in time was detected
during this period. We partition the observations by year into two data
sets referred to as \textit{2007} and \textit{2008}, respectively. Each year
is first regarded as a \textit{training set}, and each model is fit
individually to each year. The opposite year is subsequently used as a
\textit{test set} to evaluate the out-of-sample performance of each
fitted model. To account for missing days, we reinitialize the CIIR
process $\eta_t$ in the first period following each day of missing
data. This was necessary at most five times per year including the
first day of the year.

\begin{sidewaysfigure}

\includegraphics{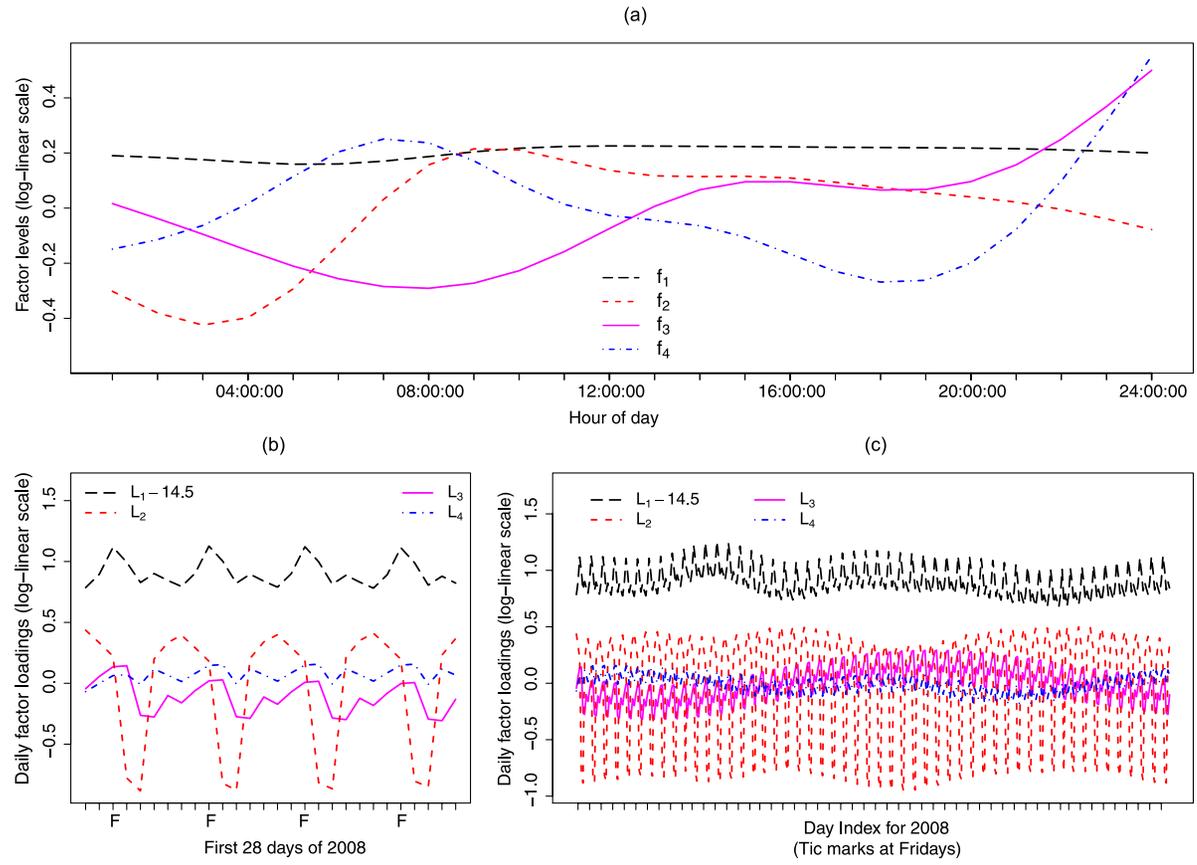}

\caption{2008 fitted \textup{(a)} factor levels $\mathbf{f}_{k}$
(log-linear scale) and [\textup{(b)} and \textup{(c)}] corresponding factor loadings $\mathbf{L}_{k
\cdot}$ (log-linear scale) for a factor model fit with constraints,
smoothing splines and $K=4$ factors. $(\mathbf{L}_{1\cdot} - 14.5)$ is
shown for easier comparison.} \label{4factor}
\end{sidewaysfigure}

We found the factor model fit with constraints, smoothing splines, and
$K = 4$ factors to be the most appropriate of the factor models
considered. The estimated factors $\mathbf{f}_k$ for 2008 are shown in
Figure \ref{4factor}(a). Each of the four factors varies smoothly over
the hours of the day via use of smoothing splines. The first factor
$\mathbf{f}_1$ is strictly positive and the least variable. It appears
to capture the mean diurnal pattern. The factor $\mathbf{f}_2$ appears
to isolate the dominant relative differences between weekdays and
weekend days. The defining feature of $\mathbf{f}_3$ and $\mathbf
{f}_4$ is the large increase late in the day, corresponding closely to
the relative increase observed on Friday and Saturday evenings.
However, $\mathbf{f}_3$ decreases in the morning, while $\mathbf
{f}_4$ increases in the morning and decreases in the late afternoon. As
$K$ increases, additional factors become increasingly more variable
over the hours of the day. Too many factors result in overfitting the
model, as the extra factors capture noise.

The corresponding daily factor loadings $\mathbf{L}$ for the first
four weeks of 2008 are shown in Figure \ref{4factor}(b). The loadings
$(\mathbf{L}_{1} -14.5)$ are shown to simplify comparisons. The much
higher loadings on $\mathbf{f}_1$ confirm its interpretation as
capturing the mean. The peaks on Fridays coincide with Friday having
the highest average number of calls, as seen in Figure \ref{calls}.
Weekdays get a positive loading on $\mathbf{f}_2$, while weekend days
get negative loading. Loadings on $\mathbf{f}_3$ are lowest on Sundays
and Mondays and loadings on $\mathbf{f}_4$ are largest on Fridays and
Saturdays. As $K$ increases, the loadings on additional factors become
increasingly close to zero. This partially mitigates the overfitting
described above. Factors with loadings close to zero have less impact
on the fitted values $\widehat{\mu}_t$. Nevertheless, they can still
reduce out-of-sample forecast performance.

\begin{figure}

\includegraphics{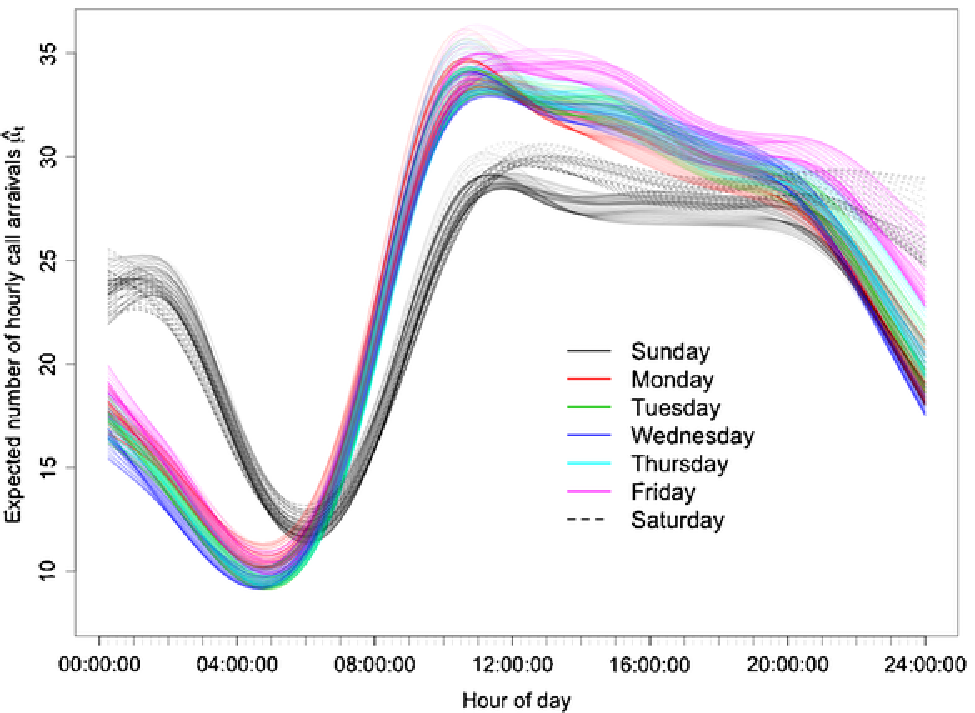}
\vspace*{-5pt}
\caption{The estimated intensity process $\hat{\bolds{\mu}}_i$,
for every day in 2008, for a factor model fit with constraints,
smoothing splines and $K=4$ factors, colored by day-of-week, and shaded
light to dark by week-of-year.} \label{4smooth}
\vspace*{-6pt}
\end{figure}

The daily factor loadings for all of 2008 are shown in Figure \ref
{4factor}(c). The relative magnitude of each loading vector with
respect to day-of-week is constant. This results from use of the
constraint matrix $\mathbf{H}^{(1)}$ in (\ref{m4}). As the loadings
vary over the days of the week, they also vary smoothly over the course
of the year, via use of the constraint matrix $\mathbf{H}^{(2)}$ and
the use of cyclic smoothing splines in estimation of $\mathbf
{B}^{(2)}$ in (\ref{m4}). The loadings on $\mathbf{f}_1$ show how the
expected number of calls per day varies over the year. The week to week
variability in the other loadings influences how the days of the week
change relative to each other over the year. Figure \ref{4smooth}
shows the estimated intensity process $\widehat{\bolds{\mu}}_i$ for
every day in 2008, shaded by day-of-week. The curves vary smoothly over
the hours of the day. The fit for each day of the week keeps the same
relative shape, but it varies smoothly over the weeks of the
year.

\begin{figure}

\includegraphics{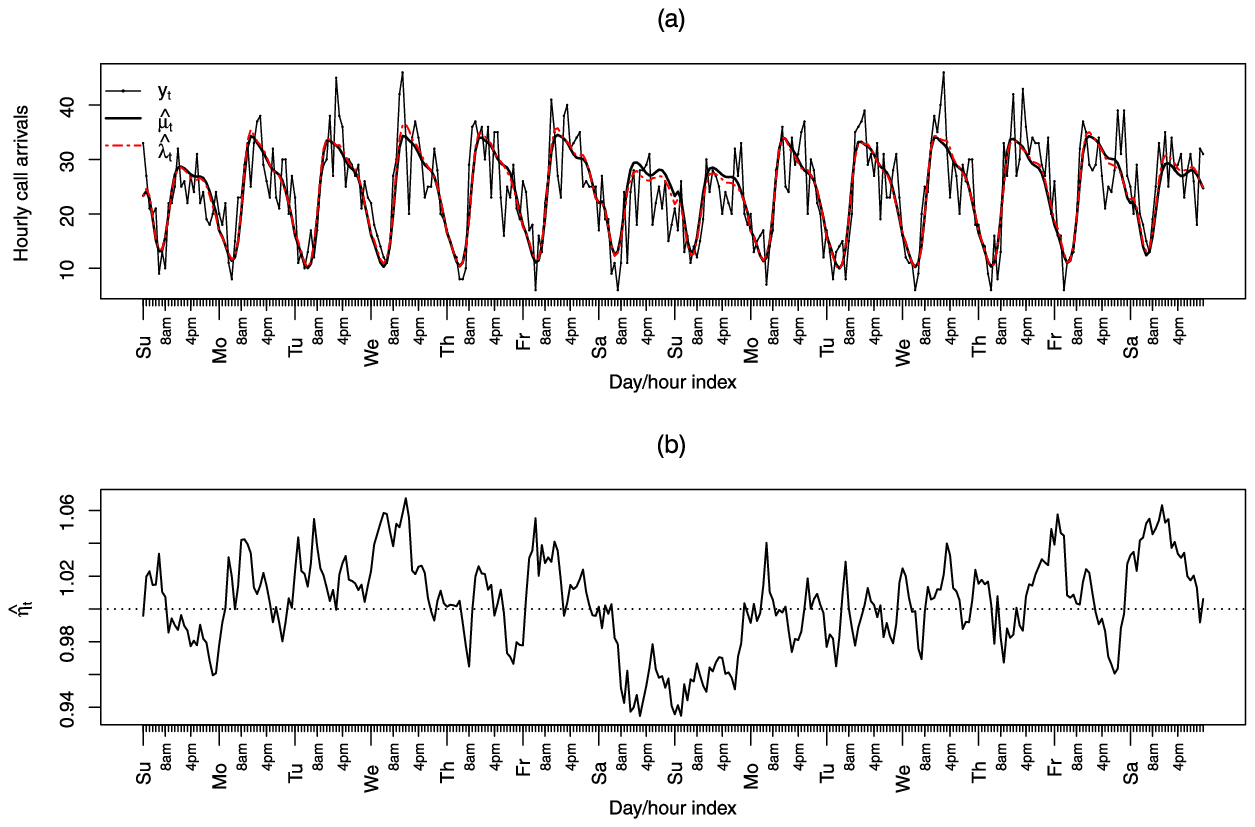}

\caption{Weeks 8 and 9 of 2007:
\textup{(a)} observed call arrivals per hour $y_t$,
fitted $K=4$ dynamic factor model $\widehat{\mu}_t$ using constraints and smoothing splines,
and factor model $\widehat{\lambda}_t$ including fitted $\operatorname{IntGARCH}(1,1)$;
\textup{(b)} the fitted conditional intensity inflation process $\widehat{\eta}_t$ from the $\operatorname{IntGARCH}(1,1)$ model.}
\label{data1}
\end{figure}

Section \ref{3.4} described incorporating time series models to
improve the short-term forecasts of a factor model. The models capture
the observed serial dependence in the multiplicative residuals from a
fitted factor model; see Figure \ref{ACF}. Parameter estimates and
approximate standard errors for the IntGARCH model are given in Supplemental material
(Table~1). A fitted factor model $\widehat{\mu}_t$ using
constraints, smoothing splines and $K=4$, as well as the factor model
including a fitted $\operatorname{IntGARCH}(1,1)$ model $\widehat{\lambda}_t$, are
also shown in Figure \ref{data1}(a),\vspace*{1pt} with the observed call arrivals
per hour for Weeks 8 and 9 of 2007. The $\widehat{\lambda}_t$ process
is mean reverting about the $\widehat{\mu}_t$ process. They are
typically close to each other, but when they differ by a larger amount,
they tend to differ for several hours at a time. The corresponding
fitted CIIR process $\widehat{\eta}_t$ is shown in Figure \ref
{data1}(b). This clearly illustrates the dependence and persistence
exhibited in Figure \ref{data1}(a). The CIIR process ranges between
$\pm$6\% during this period. With a mean of 24 calls per hour, this
range corresponds to $\widehat{\lambda}_t$ varying about $\widehat
{\mu}_t$ by about $\pm$1.5 expected calls per hour.

\subsection{Goodness of fit and model selection}\label{4.2}

To evaluate the fitted values and forecasts of the proposed models,
three types of residuals are computed: multiplicative, Pearson and
Anscombe. Their respective formulas for the Poisson distribution are
given by
\[
\widehat{r}_{M,t}=\frac{y_t}{{\widehat\lambda_t}} - 1,\qquad
\widehat{r}_{P,t}=\frac{y_t-\widehat\lambda_t}{\sqrt{\widehat
\lambda_t}},\qquad
\widehat{r}_{A,t}=\frac{({3}/{2})(y_t^{2/3}-\widehat\lambda
_t^{2/3})}{\widehat\lambda_t^{1/6}}.
\]
We refer to the root mean square error (RMSE) of each metric as RMSME,
\mbox{RMSPE} and RMSAE, respectively. The multiplicative residual is defined
as before and is a natural choice given the definition for the CIIR.
Since the variance of a Poisson random variable is equal to its mean,
the Pearson residual is quite standard. However, the Pearson residual
can be quite skewed for the Poisson distribution [cf. \citet
{mccuneld1989}, Section 2.4]. The Anscombe residual is derived as a
transformation that makes the distribution of the residuals as close to
Gaussian as possible while suitably scaling to stabilize the variance.
See \citet{pierce1986residuals} for further discussion of residuals
for generalized linear models. While the three methods always yielded
the same conclusion, we found use of the Anscombe residuals gave a~more
robust assessment of model accuracy and simplified paired comparisons
between the residuals of competing models.

The three RMSE metrics were used for both in- and out-of-sample model
comparisons. For in-sample comparisons of the factor models, we also
computed the \textit{deviance} of each fitted model $\widehat{\mu}_t$.
As a goodness-of-fit metric, deviance is derived from the logarithm of
a ratio of likelihoods. For a log likelihood function $\ell( \bolds{\mu
} | \mathbf{Y})$, it is defined as
\[
-2 \{ \ell( \bolds{\mu} = \widehat{\bolds{\mu}} | \mathbf{Y}) - \ell(
\bolds{\mu} = \mathbf{Y} | \mathbf{Y}) \},
\]
in general. For a fitted factor model, ignoring serial dependence, the
deviance corresponding to a Poisson distribution is
\[
2 \sum^n_{t=1} \{ y_t \log(y_t/\widehat{\mu}_t) - (y_t - \widehat
{\mu}_t) \},
\]
in which the first term is zero if $y_t = 0$.

\begin{figure}

\includegraphics{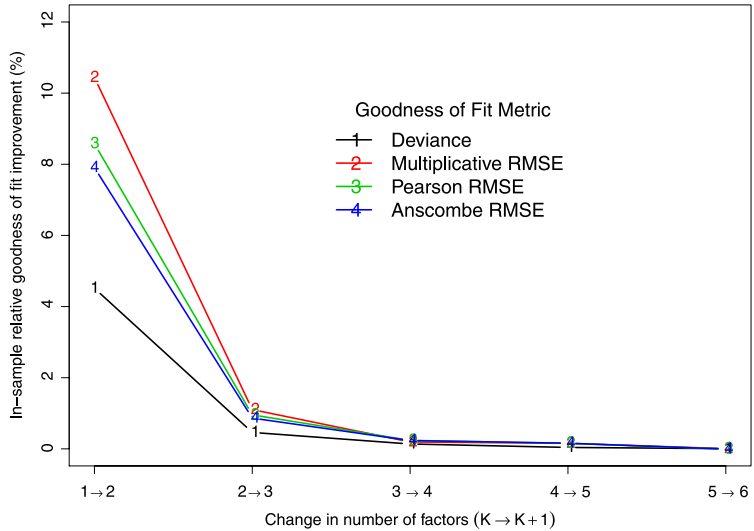}
\vspace*{-5pt}
\caption{2007 percentage in-sample relative
goodness-of-fit improvement by addition of one factor ($K \rightarrow
K+1$) for a factor model fit with constraints and smoothing splines.}
\label{deviance}
\vspace{-6pt}
\end{figure}

We compare the fitted models' relative reduction in deviance and RMSE
as we increase the number of factors $K$. Figure \ref{deviance} shows
these results for factor models fit to 2007 data with constraints and
smoothing splines. The results for other models and for 2008 were very
similar. This plot may be interpreted similarly to a \textit{scree} plot
in PCA by identifying the point at which performance tapers off and the
marginal improvement from additional factors is negligible. Under each
scenario we consistently selected $K=4$ factors through this graphical
criterion. To further justify this as a factor selection strategy, we
also consider the impact the number of factors $K$ has on out-of-sample
performance for each of the proposed models below. This approach is
straightforward, but it does not fully account for the uncertainty on
the number of factors. Bayesian estimation would require specialized
computation, but it may improve model assessment [see, e.g., \citet
{lopes2004bayesian}].

\subsection{Out-of-sample forecast performance}\label{4.3}

Out-of-sample comparisons we\-re made by fitting models to the 2007
training set and forecasting on the 2008 test set, and vice versa. To
make predictions comparable from one year to the next, we align
corresponding calendar weeks of the year, not days of the year. This
ensures that estimates for Sundays are appropriately compared to
Sundays, etc.

The first model considered was the \textit{simple prediction} (SP)
method. This simple moving average involving four observations was
defined in the\break \hyperref[sec:intro]{Introduction}. Next, the forecasts of various factor
models (FM) were considered. For $K = 1,\ldots,6,$ we evaluated the
forecasts from the FM in (\ref{m1}), the FM with constraints in (\ref
{m4}), and the FM with constraints and smoothing splines in (\ref
{m33}). Finally, for the latter FM, with $K=4$, we calculate the
implied fit from the training set with the inclusion of the CIIR
process via the various time series models defined in Section \ref
{3.4}. We compute the forecast RMSE of each model for the three
residual types, for both years.

\begin{sidewaystable}
\tablewidth=\textheight
\tablewidth=\textwidth
\caption{Root mean square multiplicative, Pearson, and
Anscombe errors for fitting model to 2007 and forecasting 2008, and
vice versa}\label{RMSFE}
\begin{tabular*}{\tablewidth}{@{\extracolsep{\fill}}lcccccccc@{}}
\hline
& & & \multicolumn{3}{c}{\textbf{2007 model, 2008 residuals}} & \multicolumn{3}{c@{}}{\textbf{2008 model, 2007 residuals}} \\[-5pt]
& & & \multicolumn{3}{c}{\hrulefill}& \multicolumn{3}{c@{}}{\hrulefill}\\
\textbf{Model}& \textbf{Constraint} & \textbf{Smoothing} & \textbf{RMSME} & \textbf{RMSPE} & \textbf{RMSAE} & \textbf{RMSME} & \textbf{RMSPE} & \textbf{RMSAE} \\
\hline
 Simple prediction & NA & NA & 0.2696 & 1.1955 & 1.1849 & 0.2661 & 1.1902 & 1.1925\\[2pt]
 Factor model, $K=1$ & No & No &  0.2722 & 1.2369 & 1.2237 & 0.2657 & 1.2183 & 1.2263\\
 Factor model, $K=2$ & No & No &  0.2721 & 1.2357 & 1.2225 & 0.2661 & 1.2197 & 1.2277\\
 Factor model, $K=3$ & No & No &  0.2727 & 1.2374 & 1.2239 & 0.2659 & 1.2182 & 1.2262\\
 Factor model, $K=4$ & No & No & 0.2729 & 1.2383 & 1.2249 & 0.2666 & 1.2206 & 1.2283\\
 Factor model, $K=5$ & No & No &  0.2732 & 1.2395 & 1.2260 & 0.2670 & 1.2220 & 1.2294\\
 Factor model, $K=6$ & No & No & 0.2733 & 1.2401 & 1.2270 & 0.2668 & 1.2217 & 1.2294\\[2pt]
 Factor model, $K=1$ & Yes & No & 0.2638 & 1.1863 & 1.1756 & 0.2575 & 1.1633 & 1.1721\\
 Factor model, $K=2$ & Yes & No & 0.2402 & 1.0938 & 1.0888 & 0.2333 & 1.0722 & 1.0875\\
 Factor model, $K=3$ & Yes & No & 0.2392 & 1.0877 & 1.0829 & 0.2324 & 1.0688 & 1.0848\\
 Factor model, $K=4$ & Yes & No & 0.2413 & 1.0945 & 1.0889 & 0.2347 & 1.0761 & 1.0912\\
 Factor model, $K=5$ & Yes & No & 0.2425 & 1.0994 & 1.0933 & 0.2363 & 1.0817 & 1.0961\\
 Factor model, $K=6$ & Yes & No & 0.2436 & 1.1051 & 1.0988 & 0.2377 & 1.0858 & 1.0999\\[2pt]
 Factor model, $K=1$ & Yes & Yes &0.2633 & 1.1837 & 1.1731 & 0.2573 & 1.1615 & 1.1703\\
 Factor model, $K=2$ & Yes & Yes & 0.2371 & 1.0844 & 1.0805 & 0.2310 & 1.0643 & 1.0803\\
 Factor model, $K=3$ & Yes & Yes & 0.2347 & 1.0744 & 1.0710 & 0.2289 & 1.0561 & 1.0728\\
 Factor model, $K=4$ & Yes & Yes & 0.2344 & 1.0730 & 1.0696 & 0.2288 & 1.0549 & 1.0715\\
 Factor model, $K=5$ & Yes & Yes & 0.2347 & 1.0740 & 1.0706 & 0.2289 & 1.0549 & 1.0714\\
 Factor model, $K=6$ & Yes & Yes & 0.2347 & 1.0739 & 1.0705 & 0.2289 & 1.0551 & 1.0716\\[2pt]
Time series and FM, $K=4$ & Yes & Yes & -- & -- & -- & -- & -- & -- \\
IntGARCH  & -- & -- & 0.2308 & 1.0571 & 1.0570 & 0.2274 & 1.0442 & 1.0580\\
IntExpGARCH & -- & -- & 0.2308 & 1.0570 & 1.0569 & 0.2274 & 1.0441 & 1.0579\\
 IntThreshGARCH & -- & -- &  0.2308 & 1.0571 & 1.0570 & 0.2275 & 1.0443 & 1.0580 \\
IntRsGARCH & -- & -- & 0.2299 & 1.0540 & 1.0554 & 0.2274 & 1.0433 & 1.0565\\
\hline
\end{tabular*}
\legend{A Yes in the constraints column implies that the factor
model was fit using the constraints outlined in Section \ref{3.2}.  A Yes in
the smoothing column indicates that the model was fit using smoothing
splines as described in Section \ref{3.3}.}
\end{sidewaystable}

The forecast results are shown in Table \ref{RMSFE}. The basic FMs did
slightly worse than the SP both years. With only one year of
observations, these FMs tend to overfit the training set data, even
with a small number of factors. The FMs with constraints give a very
significant improvement over the previous models. The forecast RMSE is
lowest at $K = 4$ for the 2007 test set, and at $K=3$ for the 2008 test
set. There was also a very large decrease between $K = 1$ and $K = 2$.
The FMs with constraints and smoothing splines offered an additional
improvement. The forecast RMSE is lowest at $K = 4$ for both test sets.
With the addition of the IntGARCH model for the CIIR process to this
model, the RMSE improved again. Application of the nonlinear time
series models instead offered only a slight improvement over the
IntGARCH model.

With only one year of training data, each FM begins to overfit with
\mbox{$K=5$} factors. Results were largely consistent regardless of the
residual used, but the Anscombe residuals were the least skewed and
allowed the simplest pairwise comparisons. Although the FMs with
constraints had superior in-sample performance, the use of smoothing
splines reduced the tendency to over-fit and resulted in improved
forecast performance. The CIIR process offered improvements in fit over
FMs alone.

We also fit each of the nonlinear time series models discussed in
Section \ref{3.4} using a FM with $K=4$. The regime switching model had the
best performance. It had the lowest RMSE for both test sets. The
exponential autoregressive and the piecewise linear threshold models
performed similarly to the IntGARCH model for both test sets. Although
the nonlinear models consistently performed better in-sample, their
out-of-sample performance was similar to the IntGARCH model.

\subsection{Queueing model simulation to approximate ambulance
operations} \label{4.4}

To comprehensively improve ambulance operations, it would be
advantageous to simultaneously model the service duration of dispatched
ambulances in addition to the demand for ambulance service.
Unfortunately, such information was not available.
We are currently working with Toronto EMS to use our improved estimates
of call arrival rates to improve staffing in their dispatch call
center. Extending our approach to a spatial-temporal forecasting model
will likely be used to help determine when \textit{and} where to deploy
ambulances.

We present a simulation study that uses a simple queueing system to
quantify the impact that improved forecasts have on staffing decisions
and relative operating costs, for the Toronto data.
The queueing model is a~simplification of ambulance operations that
ignores the spatial component. Similar queueing models have been used
frequently in EMS modeling [see \citet{swersey1994deployment}, page 173].
This goodness-of-fit measure facilitates model comparisons and a~%
similar approach may be useful in other contexts.

We use the terminology employed in the call center and queueing theory
literature throughout the section; for our application, servers are a
proxy for ambulances, callers or customers are those requiring EMS, and
a server completing service is equated to an ambulance completing
transport of a~person to a hospital, etc. As before, let $y_t$ denote
the observed number of call arrivals during hour $t$. Our experiment
examines the behavior of a simple $M/M/s$ queueing system. The arrival
rate in time period $t$ is $\lambda_t$. During this period, let $s_t$
denote the number of servers at hand. For simplicity, we assume that
the service rate $\nu$ for each server is the same, and constant over
time. Furthermore, intra-hour arrivals occur according to a Poisson
process with rate $\lambda_t$, and service times of callers are
independent and exponentially distributed with rate $\nu$.

As in Section \ref{4.3}, models are calibrated on one year of
observations and forecasts for $\lambda_t$ are made for the other
year. Each model's forecasts $\widehat{\lambda}_t$ are then used to
determine corresponding staffing levels $\widehat{s}_t$ for the system.

To facilitate comparisons of short-term forecasts, we assume that the
number of servers can be changed instantaneously at the beginning of
each period.
In practice, it is possible to adjust the number of ambulances in real
time, but not to the degree that we assume here.

Each call has an associated arrival time and service time. When a call
arrives, the caller goes immediately into service if a server is
available, otherwise it is added to the end of the queue. A common goal
in EMS is to ensure that a certain proportion of calls are reached by
an ambulance within a prespecified amount of time. We approximate this
goal by instead aiming to answer a proportion, $\theta$, of calls
immediately; this is a standard approximation in queueing applications
in many areas including EMS [\citet{kolesar1998insights}].
For each call arrival, we note whether or not the caller was served
immediately. As servers complete service, they immediately begin
serving the first caller waiting in the queue, otherwise they await new
arrivals if the queue is currently empty. One simulation replication of
the queueing system simulates all calls in the test year.


To implement the queueing system simulation, it is first necessary to
simulate arrival and service times for each caller in the forecast
period. We use the observed number of calls for each hour $y_t$ as the
number of arrivals to the system in period $t$. Since arrivals to the
system are assumed to follow a Poisson process, we determine the $y_t$
call arrival times using the well-known result that, conditional on the
total number of arrivals in the period $[t,t+1]$, the arrival times
have the same distribution as the order statistics of $y_t$ independent
Uniform($t,t+1$) random variables. We exploit this relationship to
generate the intra-hour arrival times given the observed arrival volume
$y_t$. The service times for each call are generated independently with
an Exponential($\nu$) distribution.

The final input is the initial state of the queue within the system. We
generate an initial number of callers in the queue as Poisson($y_1$),
then independently generate corresponding Exponential($\nu$) residual
service times for each of these callers.
This initialization is motivated through an infinite-server model; see,
for example, \citet{kolesar1998insights}.
Whenever there is a missing day, in either the test set or
corresponding training set period, we similarly reinitialize the state
of the queue but with $y_1$ replaced by the number of calls observed in
the first period following the missing period. These initializations
are common across the different forecasting methods to allow direct comparisons.

To evaluate forecast performance, we define a cost function and an
appropriate method for determining server levels from arrival rate
estimates. Let~$n_t$ denote the number of callers served immediately in
period $t$. The hourly cost function is given by
\[
C(n_t,y_t,s_t) = \mathrm{Pen}(n_t,y_t)+s_t,
\]
in which
\[
\mathrm{Pen}(n_t,y_t)=
\cases{
0, &\quad if $n_t \geq\theta y_t$,\cr
q(y_t-n_t), &\quad otherwise,
}
\]
$\theta\in(0,1)$ is the targeted proportion of calls served
immediately, and $q \ge0$ is the cost of not immediately serving a
customer, \textit{relative} to the cost of staffing one server for one
hour. The total cost, with respect to the hourly server cost, for the
entire forecast period is
\[
C = \sum_t C(n_t,y_t,s_t) = \sum_t \mathrm{Pen}(n_t,y_t) + \sum_t s_t.
\]
This approach, where penalties for poor service are balanced against
staffing costs, is frequently used; see, e.g., \citet
{andrews1993establishing}, \citet{harrison2005method}.

At time $t-1$, the number of call arrivals and the number served
immediately are random variables, denoted as $Y_t$ and $N_t$,
respectively. A natural objective is to choose staffing levels that
minimize the hourly expected cost as\looseness=-1
%
\begin{equation}\label{shat}
\widehat{s}_t= \operatorname{argmin}\limits_{s_t\in\mathbb{N}} E\{C(N_t,Y_t,s_t) |
\mathcal{F}_{t-1}, \mathbf{X}\},
\end{equation}
in which $Y_t$ is assumed to have a Poisson distribution with mean
equal to the arrival rate forecast $\widehat{\lambda}_t$. The staffing
levels are then a function of arrival rate forecasts, $\widehat{s}_t(\widehat
{\lambda}_t)$. We approximate this expectation numerically by randomly
generating $J$ independent realizations as $Y_{t,j} \sim
\operatorname{Poisson}(\widehat{\lambda}_t)$. Then, for each~$Y_{t,j}$ we simulate
one independent realization of $N_t$. For a fixed value of $s_t$ the
expectation is approximated by ${J}^{-1}\sum_{t=1}^J \{
\operatorname{Pen}(N_{t,j},Y_{t,j})+s_t \}$.
We found that $J=25\mbox{,}000$ provided adequate accuracy.

Independent realizations of $N_t | Y_t$ require running the queueing
system forward one hour, but this is very computationally intensive. To
approximate~$N_t | Y_t$, we use a Binomial distribution. Let
$N_{t,j}|Y_{t,j}\sim\operatorname{Binomial}\{Y_{t,j},\break g(\widehat{\lambda}_t,
s_t, \nu)\}$. The function $g$ gives the \textit{steady state}
probability that a customer is served immediately for a queueing system
with a \textit{constant} arrival rate, server level and service rate,
$\widehat{\lambda}_t, s_t$ and $\nu$, respectively. Derivation of
this function is available in any standard text on queueing theory
[e.g., \citet{gross-fundamentals}, Chapter 2].

Let $p_i$ denote the long run proportion of time such a system contains
$i$ customers and let $\rho=\lambda/(\nu s)$. Then
%
\begin{eqnarray}
g(\lambda,s,\nu)=
\cases{
\displaystyle 1-\frac{\lambda^sp_0}{s!\nu^s(1-\rho)}, &\quad if $\rho<1$,
\cr
0, &\quad  if $\rho\geq1$,
}\nonumber \\
\eqntext{\displaystyle \mbox{in which }
p_0^{-1}=\sum_{u=0}^{c-1}\frac{r^u}{u!}+\frac{r^c}{c!(1-\rho)}
\mbox{ for } \rho<1.}
\end{eqnarray}
When $\rho\geq1$, the arrival rate is faster than the net service
rate, and the system is unstable; the long run probability that a
customer is served immediately is zero. The binomial approximation
greatly reduces the computational costs and provides reasonable
results, though it tends to underestimate the true variability of $N_t
| Y_t$ due to the positive correlation in successive caller delays.

A final deliberation is needed on the removal of servers when $\widehat
{s}_t$ decreases. In our implementation, idle servers were removed
first, and, if necessary, busy servers were dropped in ascending order
with respect to remaining service time. We also considered random
selection of servers to be dropped. Doing so produced highly variable
results, and is under further study. To further simplify the
implementation, if it was necessary to drop a busy server, it was
simply discarded, along with any remaining service time for that
caller. The effect of this simplification depends on the service rate
$\nu$; our results did not appear to be sensitive to this simplification.

Simulation of the queueing system is now rather straightforward. On
each iteration $i$, we note whether each caller was served immediately
or not. Forecast performance is assessed by examining the total cost
$C^{(i)} = \sum_{t} C(n_t^{(i)}, y_t, \hat{s}_t)$ over the test
period. For both years, we performed 100 simulations over the test year
for each forecast method. To demonstrate the robustness of this
methodology, we performed the experiment for several different values
of the queuing system's parameters. Specifically, all combinations of
$q\in\{2,5,10\},\ \nu\in\{1,\frac{2}{3}\},$ and $\theta\in\{
0.8,0.9\}$ were considered, after consultation with EMS experts.

\begin{figure}

\includegraphics{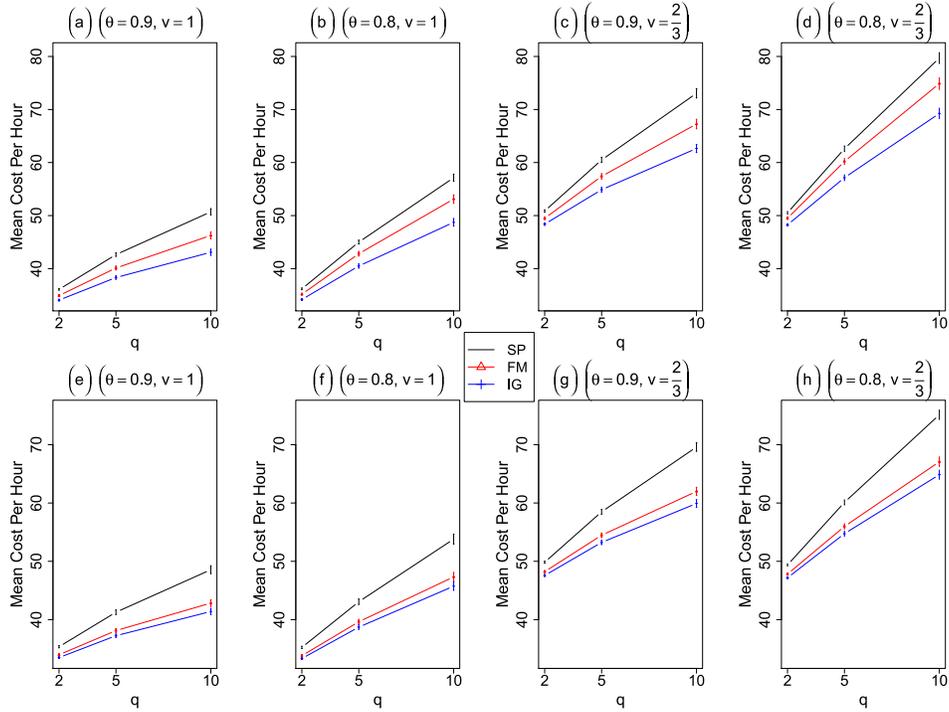}

\caption{Mean total per period cost over 100
simulations for different forecasting methods and different values of
$q$,  $\nu$ and $\theta$.  Plots \textup{(a)--(d)} use the 2008 test set and
plots \textup{(e)--(h)} use 2007 as the test set. The vertical lines represent
$\pm$1 standard deviation.}\label{totalcostfig}
\end{figure}

Results for the mean hourly cost over the 100 simulations for each
forecasting method, for each test year, are summarized in Figure \ref
{totalcostfig}. We see that the mean hourly cost is lowest for the FM
w/ IntGARCH, followed by the FM only, and finally by SP. All pairwise
differences in mean were highly significant; the smallest $t$-ratio was
80. In fact, this ordering in performance held for almost every
iteration of the queueing system, not just on average.

\begin{figure}

\includegraphics{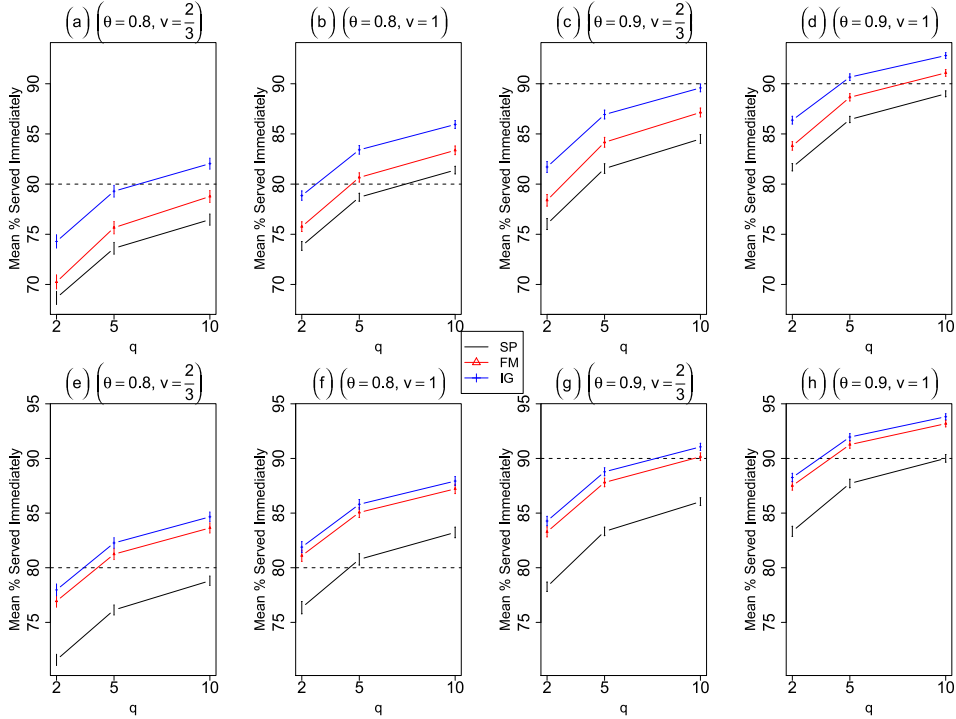}

\caption{Mean percentage served immediately for the
entire test set over 100 simulations for different forecasting methods
and different values of $q$, $\nu$ and $\theta$.  Plots \textup{(a)--(d)} use
the 2008 test set and plots \textup{(e)--(h)} use 2007 as the test set. The
vertical lines represent $\pm$1 standard deviation.}\label{psifig}
\end{figure}

The mean percentage of callers served immediately can be found in
Figure~\ref{psifig}. The total number of server hours $\sum\hat
{s}_t$ used was also recorded for each model for each set of parameter
values. A table containing the values of all these quantities can be
found in the online supplemental material. Both mean percentage served
immediately and mean hourly cost increase with $q$.
For each test year, for each level of $(q,\nu,\theta)$, $\sum_t \hat
{s}_t$ differed by between one and three thousand server-hours, for the
different models.

\section{Conclusions}\label{sec:con}

Our analysis was motivated by a data set provided by Toronto EMS.
The proposed forecasting method allows parsimonious modeling of the
dependent and nonstationary count-valued EMS call arrival process. Our
method is straightforward to implement and demonstrates substantial
improvements in forecast performance relative to simpler forecasting
methods. We measured the impact of our successive refinements to the
model, showing the merit of factor model estimation with covariates and
smoothing splines. The factor model was able to capture the
nonstationary behavior exhibited in call arrivals. Introduction of the
CIIR process allowed adaptive forecasts of deviations from this diurnal pattern.

Assessing the impact that different arrival rate forecasts can have on
call centers and related applications has received very little
attention in the literature. Our data-based simulation approach is
straightforward to implement, and was able to clearly distinguish the
effectiveness of each forecasting method. The simulation results
coincide with the out-of-sample RMSE analysis in Section \ref{4.3} and
provide a practical measure of forecast performance. Relative operating
cost is a natural metric for measuring call arrival rate forecasts, and
our implementation may easily be extended to many customized cost
functions and a wide variety of applications.

\dm{Ultimately, we seek to strengthen emergency medical service by
improving upon relevant statistical methodology. Future work will
consider inclusion of additional covariates and study of other
nonlinear time series models.
Bayesian methods which directly model count-valued observations have
desirable properties for inference and many applications, and are under
study. Spatial and spatial--temporal analysis of call arrivals will also
offer new benefits to EMS.
}


\section*{Acknowledgments}
The authors sincerely thank Toronto EMS for sharing their data, in
particular, Mr. Dave Lyons for his comments and support.

\begin{supplement}[id=suppA]
\sname{Supplement A}
\stitle{Additional tables}
\slink[doi]{10.1214/10-AOAS442SUPPA} 
\slink[url]{http://lib.stat.cmu.edu/aoas/442/supplementA.pdf}
\sdatatype{.pdf}
\sdescription{Tables 1 and 2.}
\end{supplement}

\begin{supplement}[id=suppB]
\sname{Supplement B}
\stitle{Estimation algorithms}
\slink[doi]{10.1214/10-AOAS442SUPPB} 
\slink[url]{http://lib.stat.cmu.edu/aoas/442/supplementB.R}
\sdatatype{.R}
\sdescription{\textit{R} code for estimating the models
in Section \ref{3.0} and for calculating the RMSE metrics in Section \ref{4}.}
\end{supplement}

\begin{supplement}[id=suppC]
\sname{Supplement C}
\stitle{Simulation algorithms}
\slink[doi]{10.1214/10-AOAS442SUPPC} 
\slink[url]{http://lib.stat.cmu.edu/aoas/442/supplementC.R}
\sdatatype{.R}
\sdescription{\textit{R} code for implementing the
queueing model simulation in Section \ref{4.4}.}
\end{supplement}

%

\printaddresses

\end{document}